\documentclass{ieeeaccess}
\usepackage{cite}
\usepackage{amsmath,amssymb,amsfonts}
\usepackage{graphicx}
\usepackage{textcomp}
\usepackage{tikz}
\NewSpotColorSpace{PANTONE}
\AddSpotColor{PANTONE} {PANTONE3015C} {PANTONE\SpotSpace 3015\SpotSpace C} {1 0.3 0 0.2}
\SetPageColorSpace{PANTONE}
\usepackage{algorithm}  
\usepackage{algorithmicx}  
\usepackage{algpseudocode}
\usepackage{subfigure}
\usepackage{hyperref}
\usepackage{academicons}
\usepackage{xcolor}
\usepackage{adjustbox}
\usepackage{caption, multirow,enumerate}

\usepackage{tikz,xcolor,hyperref}% Make Orcid icon
\definecolor{lime}{HTML}{A6CE39}
\DeclareRobustCommand{\orcidicon}{%
    \begin{tikzpicture}
    \draw[lime, fill=lime] (0,0) 
    circle [radius=0.16] 
    node[white] {{\fontfamily{qag}\selectfont \tiny ID}};    \draw[white, fill=white] (-0.0625,0.095) 
    circle [radius=0.007];    \end{tikzpicture}
    \hspace{-2mm}}
\foreach \x in {A, ..., Z}{%
    \expandafter\xdef\csname orcid\x\endcsname{\noexpand\href{https://orcid.org/\csname orcidauthor\x\endcsname}{\noexpand\orcidicon}}
    }
\newcommand{\vct}[1]{\boldsymbol{\mathbf{#1}}}
\def\BibTeX{{\rm B\kern-.05em{\sc i\kern-.025em b}\kern-.08em
T\kern-.1667em\lower.7ex\hbox{E}\kern-.125emX}}

\begin{document}
\history{Received 11 March 2024, accepted 15 May 2024, date of publication 1 July 2024, date of current version 26 July 2024.}
\doi{10.1109/ACCESS.2024.3421281}

\title{Toward Improving Synthetic Audio Spoofing Detection Robustness via Meta-learning and Disentangled Training with Adversarial Examples}

\author{\uppercase{Zhenyu Wang}\orcidA{},
\uppercase{John H. L. Hansen}\orcidB{},
\IEEEmembership{Fellow, IEEE}}

%\thanks{The associate editor coordinating the review of this manuscript and approving it for publication was Antonio J. R. Neves\orcidC{}.}

\address{Center for Robust Speech Systems (CRSS), University of Texas at Dallas, TX 75080 USA.}

\tfootnote{This project was supported by the University of Texas at Dallas from the Distinguished University Chair in Telecommunications Engineering held by J. H. L. Hansen.}

\markboth
{Z. Wang, J. H. L. Hansen: Toward Improving Synthetic Audio Spoofing Detection Robustness}
{Z. Wang, J. H. L. Hansen: Toward Improving Synthetic Audio Spoofing Detection Robustness}

\corresp{Corresponding authors: John H.L. Hansen (e-mail: john.hansen@utdallas.edu).}

\begin{abstract}
Advances in automatic speaker verification (ASV) promote research into the formulation of spoofing detection systems for real-world applications. The performance of ASV systems can be degraded severely by multiple types of spoofing attacks, namely, synthetic speech (SS), voice conversion (VC), replay, twins and impersonation, especially in the case of unseen synthetic spoofing attacks. A reliable and robust spoofing detection system can act as a security gate to filter out spoofing attacks instead of having them reach the ASV system. A weighted additive angular margin loss is proposed to address the data imbalance issue, and different margins has been assigned to improve generalization to unseen spoofing attacks in this study. Meanwhile, we incorporate a meta-learning loss function to optimize differences between the embeddings of support versus query set in order to learn a spoofing-category-independent embedding space for utterances. Furthermore, we craft adversarial examples by adding imperceptible perturbations to spoofing speech as a data augmentation strategy, then we use an auxiliary batch normalization (BN) to guarantee that corresponding normalization statistics are performed exclusively on the adversarial examples. Additionally, A simple attention module is integrated into the residual block to refine the feature extraction process. Evaluation results on the Logical Access (LA) track of the ASVspoof 2019 corpus provides confirmation of our proposed approaches' effectiveness in terms of a pooled EER of 0.87\%, and a min t-DCF of 0.0277. These advancements offer effective options to reduce the impact of spoofing attacks on voice recognition/authentication systems.  
\end{abstract}


\titlepgskip=-15pt

\maketitle

\section{Introduction}
 
\label{sec:introduction}
\PARstart{I}{n} recent years, ASV has been used extensively for personal biometric authentication. An ASV system aims to verify an identity claim of an individual from their voice characteristics \cite{hansen2015speaker}. Spoofed voice attacks involve an attacker who masquerades as the target speaker to gain access into the ASV system \cite{evans2013spoofing,ergunay2015vulnerability} for use of resources, services or devices. In most cases, the zero-effort imposters can be easily caught by a general ASV system, but more sophisticated spoofing attacks pose a significant threat to system robustness and credibility \cite{kamble2020advances}. With easy access to biometric information of personal voices, spoofing attacks are inevitable \cite{galbally2014biometric}. Such a potential system security breach represents a key reliability concern of ASV systems. To address this, an audio spoofing detection system generates countermeasure scores for each audio sample to distinguish between genuine (bona-fide) and spoofed speech, which allows for deployment of the ASV system into real-world situations where diverse audio spoofing attacks could occur. 
\vspace{-0.2cm}

Since 2015 \cite{wu2015asvspoof,kinnunen2017asvspoof,todisco2019asvspoof,yamagishi2021asvspoof}, the ASVspoof community has been at the forefront of anti-spoofing research with a series of biannual challenges. Their aims are to foster progress in development of audio spoofing detection to protect ASV systems from manipulation. Existing audio spoofing detection systems have been proposed to address two different mainstream use case scenarios: logical access (LA) and physical access (PA), which involves three major forms of spoofing attack, namely synthetic, converted, and replayed speech. Spoofing attacks on the physical access track are direct attacks at the transmission stage, where genuine audio samples are represented by a replay device to microphone input of the ASV system\cite{wu2014study}. With continuous progress in speech synthesis \cite{masuko1999security} and voice conversion \cite{pellom1999experimental}, such advanced techniques are able to impersonate a target speaker's voice, compromising ASV reliability. Logical access attacks, generated by the latest speech synthesis and voice conversion technologies, can be more challenging and perceptually indistinguishable from genuine speech.

Such spoofed speech generated by different attacking algorithms contains artefacts, which reside in specific sub-bands or temporal segments \cite{tak2020explainability,sahidullah2015comparison,sriskandaraja2016investigation,yang2019significance,garg2019subband,chettri2020subband,tak2020spoofing}. Specifically, artefacts serve as indicative cues to distinguish genuine speech from spoofed speech. Additionally, artefacts present in different attacks tend to be heterogeneous, which depend on the specific spoofing algorithm employed. Reliable detection often relies upon the ensemble system with multiple subsystems tuned to capture specific forms of artefacts. Here, we seek to develop a single system that delivers reliable detection performances across a spectrum of diverse spoofing attacks.

As in many related fields of speech processing, a growing number of researchers are adopting end-to-end model architectures that operate directly upon raw speech waveforms \cite{hajibabaei2018unified,nagrani2020voxceleb,jung2018avoiding,ravanelli2018learning}, which bypass limitations introduced by the utilization of knowledge-based, hand-crafted acoustic features, (e.g., Mel-frequency cepstral coefficients, and Melfilterbank energy features\cite{variani2014deep,snyder2018x,okabe2018attentive,jung2019spatial}). Following this trend, RawNet2 \cite{Jung2020ImprovedRW}, combined with the merits of RawNet1 \cite{jung2019rawnet}, takes in raw waveforms and tends to yield more discriminative representations compared to traditional spoofing detection solutions. To learn a meaningful filterbank structure, the first layer of RawNet2 is the same as that of SincNet \cite{ravanelli2018speaker,ravanelli2018learning}, which implements band-pass filters based on parametrized sinc functions. The upper layers are comprised of residual blocks \cite{he2016deep} to extract frame-level representations, and the GRU \cite{chung2014empirical} layer serves to aggregate utterance-level representations. Here, filter-wise feature map scaling (FMS) \cite{Jung2020ImprovedRW} is employed as an attention mechanism to derive more discriminative representations.

To further enhance the model's representation ability to construct informative features, the Squeeze-and-Excitation (SE) component has been extensively used in residual blocks, which recalibrates channel-wise feature maps by modelling the inter-dependencies between each channel \cite{hu2018squeeze}. Given an intermediate feature map in a residual block, the convolutional block attention module (CBAM) \cite{woo2018cbam} sequentially infers attention maps along the channel and spectral-temporal dimensions, and then attention maps are used to refine the input features. In contrast to channel-wise and spatial-wise attention modules, a simple attention module (SimAM) \cite{yang2021simam} infers 3-dimensional attention weights for adaptive feature refinement. Inspired by neuroscience theories, they propose to optimize an energy function to attain the importance of each neuron. Attention modules noted here represent general plug-and-play modules, which can be injected into each residual block of any feed-forward convolutional neural network (CNN) architecture seamlessly with negligible additional parameters and is also end-to-end trainable along with CNNs.

In most cases, the spoofing detection classifier is trained using a cross-entropy loss with softmax (denoted by CE-Softmax loss). A reliable spoofing detection model should aggregate embeddings from the same identity and separate clusters for different identities. However, the spoofing detection model optimized by Softmax loss is not generalizable enough, and performance degradation is observed when evaluated on unseen spoofing attacks. As in some speaker verification tasks \cite{zhang2016end,zhang2017end,sang2022self,sang2021deaan}, the end-to-end system is able to learn discriminative representations directly, however, it is time-consuming for training and requires complex data preparation (e.g., semi-hard example mining). To address this issue with negligible computational overhead, margin-based losses such as angular softmax loss (denoted by A-Softmax loss) \cite{liu2017sphereface}, additive margin softmax loss (denoted by AM-Softmax loss) \cite{wang2018additive,wang2018additive}, and additive angular margin loss (denoted by AAM-Softmax loss) \cite{deng2019arcface}, can be considered to encourage intra-class compactness and inter-class segregation. Previous research has investigated the impact of margin-based losses for speaker embedding learning \cite{sang2020open,sang2022multi,sang2023improving}, It has been proven that margin serves as a vital factor in discriminative embeddings learning and leads to a significant overall performance improvement \cite{cai2018exploring}.

Due to the continuing evolution of voice conversion and speech synthesis techniques, a growing number of emerging unseen spoofing attacks poses a great threat to the reliability of spoofing detection systems. The generalization capability of existing solutions could be subject to a limited variety of known attacks. Meta-learning has recently become one research hotspot in deep-learning-based approaches. Several novel meta-learning approaches \cite{ko2020prototypical,kye2020meta,wang2019centroid} propose to learn a shared metric space between the embeddings of unseen examples from a test set, and known classes in the training set. Ko et al. \cite{ko2020prototypical} employed prototypical networks (PN), a typical meta-learning architecture, to enhance the discriminative power of the speaker embedding extractor. While episodic optimization could be insufficient to obtain the optimal embedding distribution for unseen classes, Kye et al.\cite{kye2020meta} perform global classification for each sample within every episode. By combining two learning schemes, significant improvement is observed for short-duration utterance speaker recognition. With consideration of unseen samples in the test set during the training phase, the model achieves a consistent framework across train and test, which boosts discriminative power for unseen samples.

\begin{figure*}[htbp]
\centering
\includegraphics[scale=0.625,angle=270]{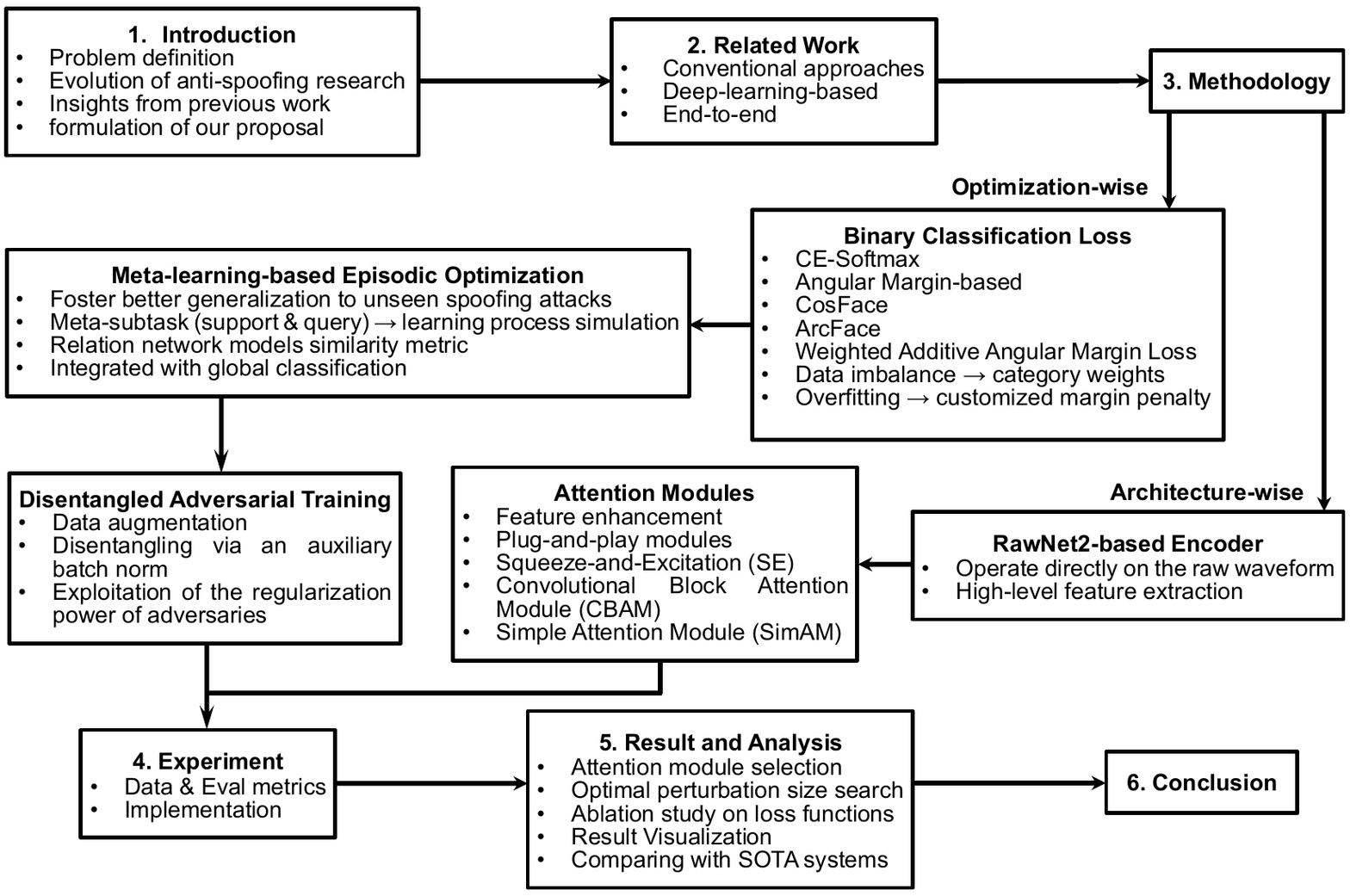}
\caption{Study overview.}
\label{fig:overview}
\vspace{-3.5ex}
\end{figure*}

Deep-learning-based approaches always require a large amount of data to tune model parameters during training, where the spoofing detection accuracy and model robustness can be subject to training data size. Data augmentation is a commonly-used method to obtain additional synthetically modified data. Adversarial examples can also be obtained by attacking model vulnerability, which has been adopted as a free resource for model training in different tasks \cite{kurakin2016adversarial} \cite{madry2017towards} \cite{pang2020boosting}. Adversaries are crafted by adding imperceptible perturbations to original training data, and the modified data is used to mislead a well-trained model \cite{carlini2018audio}. Adversarial examples are commonly viewed as a threat to neural network models, which behave in a similar manner to spoofing attacks. Inspired by this, adversaries can be treated as additional training examples to boost spoofing detection performance if harnessed in the proper manner. Although there probably exists a trade-off between model accuracy and robustness to adversarial perturbations \cite{freelunch}. Unexpected benefits can be observed when adversarial examples are involved in the model training (e.g., interpretable feature representations that align well with salient data characteristics \cite{freelunch} and improved robustness to corruptions concentrated in the high-frequency domain \cite{yin2019fourier}). Considering different data distributions between original training data and adversaries, Xie et al. proposed an auxiliary batch normalization (BN) to disentangle model training for accurate statistics estimation \cite{xie2020adversarial}.

In this study, we begin with a variant of RawNet2 \cite{Jung2020ImprovedRW} as our backbone model architecture. We develop an end-to-end robust spoofing detection system to reliably detect spoofing attacks on the LA track of the ASVspoof2019 corpus without score-level ensembles. We inject three different attention modules (e.g., SE, CBAM, SimAM) into each residual block, respectively, to enhance model representation ability. Zhang et al \cite{zhang2021one} proposed a one-class learning to improve detection performance on unknown synthetic spoofing attacks, which results from over-fitting of known attacks. Based on the AAM-Softmax loss \cite{deng2019arcface}, we assign different weights and margins to each class (e.g., genuine and spoofed) for alleviating the unbalanced data and over-fitting problem. To mitigating the adverse impact of unseen spoofing attacks, we further adopt the meta-learning loss to adaptively learn a shared metric space between unseen samples and known attacks. The relation network \cite{sung2018learning} is employed to compare samples in the support and query sets, where an additional neural network serves to parameterize the comparison metric. Previous research has explored taking adversarial examples as augmentation data to improve an attention-based keyword spotting system \cite{wang2019adversarial}. Here, Our interest lands on disentangled learning with adversarial examples to enhance system robustness, such that the complementarity across original training data and adversaries could be fully exploited. We hypothesize that the AAM-Softmax loss for global classification, the meta-learning episodic loss, and the loss from adversarial examples exhibit specific unexpected benefits during model training, thus bringing out an overall joint optimization as a powerful ensemble method for discriminative representations learning. The main contributions of this research study are:

1. Investigated three extensions to the RawNet2-based model, and analyzed the effectiveness of each attention module in improving model performance.

2. A weighted AAM-Softmax loss is employed for binary classification to encourage intra-class compactness and inter-class separability in the embedding space.

3. proposed a meta-learning framework to enhance model generalization capability to unseen spoofing attacks, and integrated episodic and global classification to encourage discriminative embedding learning.

4. Adversarial examples are treated as additional training samples, with an auxiliary BN used for adversaries to perform disentangled training.

5. Joint optimization with weighted AAM-Softmax loss, meta-learning loss, and adversarial loss are performed to boost the entire spoofing detection system performance for detection of LA-based spoofing attacks.

The rest of the paper is organized as follows. Sec. \ref{related} investigates conventional and existing state-of-the-art spoofing detection approaches. Sec. \ref{method} details each component of the proposed spoofing detection framework. Sec. \ref{experiment} comprises the specifics of data, evaluation metrics, and experimental configurations. Sec. \ref{res} presents experimental results and corresponding analysis of observations. Lastly, conclusions are drawn in Sec. \ref{conclusion}. The overview of this study is presented in Fig. \ref{fig:overview}.

%In light of the recent successes of 
\section{related work}
\label{related}
This section presents a detailed investigation of existing state-of-the-art countermeasures for audio synthetic spoofing detection. The countermeasures are broadly classified into three categories: conventional handcrafted features with machine learning classifiers, enhanced deep learning approaches, and state-of-the-art end-to-end approaches.
\subsection{conventional approaches}
Researchers in the spoofing detection community have worked on finding handcrafted features that reflect artefacts based on phase spectrum, magnitude spectrum, pitch, group delay, etc., to distinguish between spoofed and genuine speech \cite{de2012synthetic,todisco2017constant,alam2015development,wu2013synthetic,de2011detection,saratxaga2016synthetic}. Since feature extraction and classifiers are two main components of spoofing detection systems, the Gaussian mixture
model (GMM), its variants, and support vector machine (SVM) classifiers \cite{de2012synthetic,alam2015development,liu2015simultaneous,alegre2013new,wu2012detecting} have been extensively explored for synthetic spoofing detection. However, it has been shown that efforts on complex machine-learning-based classifiers are less effective than crafting informative features\cite{kamble2017novel}.

The Constant-Q Cepstral Coefficients (CQCC) \cite{todisco2017constant} are extracted with the constant-Q transform (CQT), which captures manipulation artefacts that are indicative of spoofing attacks. Patel et al. proposed a combination of cochlear filter cepstral coefficients (CFCC) and change in instantaneous frequency (IF) (i.e., CFCCIF) to detect genuine versus spoofed speech \cite{patel2015combining}. Additionally, improved classification performance was observed when CFCCIF was combined with Mel frequency cepstral coefficients (MFCC) \cite{zheng2001comparison} features. For other effective features, the high-dimensional magnitude-based features (i.e., log magnitude spectrum, and residual log magnitude spectrum) and phase-based features (i.e., group delay function, modified group delay function, baseband phase difference, pitch synchronous phase, instantaneous frequency derivative) have been introduced in \cite{xiao2015spoofing}.

Artefacts from synthetic speech reside in different subbands, therefore, subband processing is explored to extract discriminative features such as linear frequency cepstral coefficients (LFCC) \cite{sahidullah2015comparison}, energy separation algorithm instantaneous frequency cepstral coefficients (ESA-IFCC) \cite{kamble2017novel}, and constant-Q statistics-plus-principal information coefficient (CQSPIC) \cite{yang2018feature}. Kaavya et al. proposed another subband processing approach to perform a hierarchical scattering decomposition through a wavelet filterbank, then the absolute values of the filter outputs are used to yield a scalogram \cite{sriskandaraja2016front}.

Previous studies in image processing have extensively explored the concept of texture. It has been found that texture descriptors such as local binary patterns (LBP) and local ternary patterns (LTP) are effective for image classification tasks. Next, a novel countermeasure based on the analysis of sequential acoustic feature vectors using Local Binary Patterns (LBPs) was presented to detect LA attacks\cite{alegre2013new}. \cite{saratxaga2016synthetic} also employed relative phase shift features and MGDF-based features to detect synthesized/converted speech. LBPs and MGDF \cite{liu2015simultaneous} are less successful at differentiating between genuine and spoofed samples because they are susceptible to noise, which generates patterns that are similar for both classes.

\subsection{deep-learning-based approaches}
Recent efforts have witnessed a rise in utilization of deep-learning-based methods to detect synthesized/converted spoofing attacks. Alzantot et al. \cite{alzantot2019deep} built three variants of ResNet \cite{he2016deep} that ingested different feature representations, namely, MFCC, log-magnitude STFT, and CQCC. The fusion of three variants of ResNet (i.e., MFCC-ResNet, CQCC-ResNet, and Spec-ResNet) has outperformed the spoofing detection baseline methods (i.e., LFCC-GMM, CQCC-GMM). Wang et al. \cite{wang2019adversarial} used a 135-layer deep dense convolutional network to detect voice transformation spoofing. Similarly, Lai et al. \cite{lai2019assert} adopted two low-level acoustic features, namely, log power magnitude spectra (logspec) and CQCC as input, where the DNN models hinged on variants of all the Squeeze-Excitation (SE) network and residual networks were trained to detect spoofed speech. In \cite{villalba2015spoofing}, spectral log-filter-bank and relative phase shift features were taken as input to train DNN classifiers for synthetic spoofing detection. A five-layer DNN classifier with a novel human log-likelihoods (HLL) scoring method was proposed, which was mathematically proven to be more suitable for synthetic spoofing detection than the classical LLR scoring method \cite{yu2017spoofing}.

Concerning feature engineering, it was found that utilizing the DNN-based model as a pattern classifier was less effective than using it for representation learning followed by traditional machine learning classifiers (i.e., GMM or SVM as the classifier). In \cite{chen2015robust}, a spoofing-discriminant network was used to extract the representative spoofing vector (s-vector) at the utterance level. Next, the Mahalanobis distance, along with normalization, was applied to the computed s-vector for LA attack detection. In \cite{alam2016spoofing}, bottleneck features with frame-level posteriors were extracted by the DNN-based model, followed by a standard GMM classifier built with acoustic-level features and bottleneck features. In \cite{gomez2019light}, a light convolutional gated recurrent neural network was used to extract utterance-level representations, later with extracted deep features, back-end classifiers (i.e. linear discriminant analysis (LDA), and its probabilistic version (PLDA), and SVM) performed the final genuine/spoofed classification). A similar approach has been proposed to learn spoofing identity representations \cite{qian2016deep}, where DNN-based frame-level features and RNN-based sequence-level features were incorporated in model training (i.e. LDA, gaussian density function (GDF), and SVM) for spoofing detection. Despite the extra computation costs introduced by feature engineering, deep-learning-based methods deliver better classification performances than traditional methods.

\subsection{end-to-end approaches}
Today, end-to-end approaches have achieved state-of-the-art performance in a variety of audio processing applications \cite{zhang2017advanced,heittola2018machine}. Bypassing complex feature engineering, the end-to-end framework takes raw waveforms as input for representation learning and yields corresponding classification decisions, which encapsulate pre-processing and post-processing components within a single network \cite{tokozume2017learning,chiu2018state}. Muckenhirn developed a convolutional neural network-based approach to learn features and then built a classifier in an end-to-end manner \cite{muckenhirn2017end}. A joint architecture called convolutional Long-Short Term Memory (LSTM) neural network (CLDNN) with raw waveform front-ends was proposed for spoofing detection \cite{dinkel2017end,dinkel2018investigating}. In the literature \cite{tak2021end}, an end-to-end system based on a variant of RawNet2 encoder \cite{Jung2020ImprovedRW} and spectro-temporal graph attention networks \cite{velickovic2017graph} was used to learn the relationship between cues spanning different sub-bands and temporal segments. Jung et al. developed an end-to-end architecture incorporated with a novel heterogeneous stacking graph attention layer, followed by a new max graph operation and readout scheme, to facilitate the concurrent modelling of temporal-spectral graph attention for improved spoofing detection \cite{jung2022aasist}. Following previous work \cite{ge2021partially} based on a variant \cite{xu2019pc} of differentiable architecture search \cite{liu2018darts}, Ge et al. explored how to learn automatically the network architecture towards a spoofing detection solution \cite{ge2021raw}. End-to-end approaches represent a new direction of anti-spoofing study.

\section{Methodology}

This section describes each of the relevant components for building our proposed synthetic spoofing detection architecture. This comprises the encoder for general representation learning, attention modules for feature enhancement, and three specific optimization/training schemes to improve model accuracy, generalization ability to unseen attacks, and robustness.
\label{method}
\subsection{RawNet2-based Encoder}
\label{sec:encoder}
Instead of using hand-crafted features as inputs \cite{tak2021graph}, the Rawnet2-based model operates directly upon the raw waveform without preprocessing techniques \cite{hua2021towards,tak2021end}. A variant of the RawNet2 model was introduced in \cite{jung2019rawnet} for the speaker embedding learning and applied subsequently for building spoofing detection systems \cite{tak2021end,tak2021endsp}. Here, we adopt that model to extract high-level representations $F \in \mathbb{R}^{C\times S\times T}$ (C, S, and T are the number of channels, spectral bins, and the temporal sequence length, respectively) from raw waveforms. According to the literature \cite{tak2021endsp,ravanelli2018speaker,jung2019rawnet}, approaches equipped with a bank of sinc filters show superior effectiveness in terms of both convergence stability and performance. Therefore, a sinc convolution layer is employed for front-end feature learning. The sinc layer transforms the raw waveform in the time domain using a set of parameterized sinc functions that are analogous to rectangular band-pass filters \cite{quatieri2002discrete,Deller1993DiscreteTimePO}. Each filter within the filterbank possesses its center frequencies based on a mel-scale. Cut-in and cutoff frequencies are fixed to alleviate over-fitting to training data due to training data sparsity or rather limited genres of different spoofing attacks (only 6 for the training and development partitions from the ASVspoof 2019 LA database).

The output of each filter is treated as a spectral bin, subsequently, the output of the sinc layer is transformed into a 2-dim time-frequency representation by adding a channel dimension. The result is fed into stacked 2-dim residual blocks \cite{he2016deep} with pre-activation \cite{he2016identity} for high-level feature learning. Each residual block is comprised of a batch normalization layer \cite{ioffe2015batch}, a 2-dim convolution layer, scaled exponential linear units (SeLU) \cite{klambauer2017self}, and a max pooling layer for down-sampling. The specifics of our model configuration are summarized in Tab. \ref{table:model}.

\begin{table}[ht]
\renewcommand{\arraystretch}{1.3}
\captionsetup{font={footnotesize}}
 \caption{Model configuration}
  %\label{data_stat}
 \begin{adjustbox}{width=1.\columnwidth, center}
 \centering

 \vspace{-1.5ex}
 \begin{tabular}{ccc}
 \hline
 \hline 
 \multicolumn{1}{c}{Layer} & Input: 64600 samples & Output shape \\
 \hline 
 
 \multirow{4}{*}{Sinc Layer} & Conv-1D(129,1,70) & (70,64472) \\ 
 %\cline{2-4}
								 & add channel (TF representation) & (1,70,64472) \\
 %\cline{2-4}
& Maxpool-2D(3) & (1,23,21490) \\
& BN \& SeLU & \\ 
 %\cline{2-4}
 \hline
 \multirow{5}{*}{Res block $\times$ 2} & Conv-2D(32,(2,3),(1,1),1) & \multirow{4}{*}{(32,23,2387)} \\ 
 %\cline{2-4}
								 & BN\&SeLU & \\
 %\cline{2-4}
& Conv-2D(32,(2,3),(0,1),1) & \\
%& SE module ($reduction=8$) \\
& Maxpool-2D((1,3))& \\
 %\cline{2-4}
 \hline
 \multirow{5}{*}{Res block $\times$ 4} & Conv-2D(64,(2,3),(1,1),1) & \multirow{4}{*}{(64,23,29)} \\ 
 %\cline{2-4}
								 & BN\&SeLU & \\
 %\cline{2-4}
& Conv-2D(64,(2,3),(0,1),1) & \\
%& SE module ($reduction=8$) \\
& Maxpool-2D((1,3))& \\
 %\cline{2-4}
 \hline
AdaptivePooling & AdaptiveAvgPool2d((1,29))& (64,1,29)\\
 \hline
%GRU & GRU($input\_size=64,hidden\_size=64$) & (64) \\
GRU & GRU(64) & (64,) \\
 \hline
Output & FC(2) & (2,) \\
 \hline
 \hline 
 \end{tabular}%
 \label{table:model}
 \end{adjustbox}
  \vspace{-1.5ex}
\end{table}%

\subsection{Attention modules}
The fundamental building block of convolutional neural networks (CNNs) serve as the convolution operator, allowing networks to learn informative features by combining spatial and channel-wise information within the local receptive fields at each layer. Plug-and-play attention modules \cite{hu2018squeeze,woo2018cbam,yang2020gated} as an effective component can refine the intermediate feature maps within a CNN block, so as to boost the model capacity. Researchers are of interest to formulate effective attention modules for feature enhancement, which enable networks to improve the quality of channel-wise or spatial encoding throughout the feature hierarchy.
\subsubsection{Squeeze-and-Excitation}
Squeeze-and-Excitation (SE) module can be integrated into residual blocks for learning informative representations by the insertion after a non-linearity following a convolution \cite{hu2018squeeze}. The module as a computational unit is comprised of two fully connected layers to learn the importance of each channel, which is built on transforming by first compressing and then expanding the full average channel vector. Given the intermediate feature map $\vct{x} \in \mathbb{R}^{C \times S \times{T}}$ of the Residual block as input, the SE module first calculates the channel-wise mean statistics $\vct{e} \in \mathbb{R}^C$. Here, the $c$-th element of $\vct{e}$ is
 \vspace{-1ex}
\begin{eqnarray}
\vct{e}_c=\frac{1}{S \times T}\sum_{i=1}^{S}\sum_{j=1}^{T}\vct{x}_{c,i,j},
\end{eqnarray}
where $C, S$, and $T$ represent channel, frequency, and time dimensions. The SE module then scales this channel-wise mean by two fully connected layers to obtain the channel-wise attention weights $s$ of the various channels:
\begin{eqnarray}
\vct{s}=\sigma(\mathbf{W}_2 f(\mathbf{W}_1 \vct{e} + \vct{b}_1) + \vct{b}_2),
\end{eqnarray}
where $W$ and $\vct{b}$ denote the weight and bias of a linear layer. Also, $f(\cdot)$ is the activate function of the rectified linear unit (ReLU), and $\sigma(\cdot)$ is the sigmoid function.

\subsubsection{Convolutional Block Attention Module}

The convolutional block attention module (CBAM) \cite{woo2018cbam} extends channel-wise attention into two separate dimensions, referred to as the channel and spatial (frequency-temporal) attention modules. Next, the input feature maps are multiplied by attention maps for adaptive feature refinement. With the merits of a lightweight and effective module, the CBAM can be integrated into any CNN-based architecture, which has previously been successfully applied for speaker verification \cite{yadav2020frequency}. Given the input feature map $\vct{x} \in \mathbb{R}^{C \times S \times{T}}$, the overall attention process sequentially infers a 1-dim channel attention map $\mathbf{M}_\mathbf{c} \in \mathbb{R}^{C \times 1 \times{1}}$ and a 2-dim frequency-temporal attention map $\mathbf{M}_\mathbf{ft} \in \mathbb{R}^{1 \times S \times{T}}$. The feature refinement process is formulated as,
\begin{align}
\begin{split}
&\vct{x}^{\prime}=\mathbf{M}_\mathbf{c}(\vct{x}) \otimes \vct{x},\\
&\vct{x}^{\prime\prime}=\mathbf{M}_\mathbf{ft}(\vct{x}^\prime)\otimes \vct{x}^\prime,
\end{split}
\end{align}
where $\otimes$ denotes element-wise multiplication. The final refined
output $\vct{x}^{\prime\prime}$ is obtained by broadcasting the attention values (i.e., $\mathbf{M}_\mathbf{c}$ and $\mathbf{M}_\mathbf{ft}$) along with the frequency-temporal and channel dimensions accordingly.

\subsubsection{Simple Attention Module (SimAM)}
\label{sec:simam}
Inspired by attention mechanisms in the human brain based on certain well-known neuroscience theories \cite{lin2020context}, the simple attention module (SimAM) \cite{yang2021simam} is proposed to optimize an energy function for encapsulating the relevance of each neuron.
The parameter-free simple attention module (SimAM) has proven to be flexible and effective in enhancing the learning capabilities of convolution networks with negligible computational costs \cite{yang2021simam}, and subsequently applied in speaker verification \cite{qin2022simple}.
By optimizing an energy function to capture the significance of each neuron, it generates 3-dim attention weights for the feature map in a convolution layer.
 \vspace{-1ex}
\begin{align}
e_t(W_t,b_t,\vct{y},x_i) = (y_t-\hat{t})^2+\frac{1}{M-1}\sum_{i=1}^{M-1}(y_o-\hat{x}_i)^2.\label{eq:1}
\end{align}

Given the feature map $\vct{x}\in\mathbb{R}^{C\times S\times T}$ in a single channel, $t$ denotes the target neuron. $x_i$ is other neurons, where i is the index over the frequency-temporal domain and $M=S\times T$ is the number of neurons for each channel. Here, $\hat{t}=W_tt+b_t$ and $\hat{x}_i=W_tx_i+b_t$ are linear transforms for $t$ and $x_i$. Eq. \ref{eq:1} obtains its minimal value when $\hat{t}=y_o$ and $\hat{x}_i=y_t$. Considering $y_o$ and $y_t$ as two distinct values, for simplicity, binary labels (i.e., 1 and -1) are assigned to $y_o$ and $y_t$ in the final energy function with a regularizer,
 \vspace{-1ex}
\begin{align}
\begin{split}
&e_t(W_t,b_t,\vct{y},x_i) = \frac{1}{M-1}\sum_{i=1}^{M-1}(-1-(W_tx_i+b_t))^2 \\
&+(1-(W_tt+b_t))^2+\lambda{W_t}^2 \label{eq:2}.
\end{split}
\end{align}

There are extensive computational resources needed to optimize each of the neuron's attention weights using a general optimizer such as SGD. Fortunately, a closed-form solution can be leveraged to derive the transform's weight $W_t$ and bias $b_t$ with optimal energy. Specifically, the minimal energy of a neuron $x$ in an input feature map $\vct{x}\in \mathbb{R}^{C\times H \times W}$ is formulated as:
\begin{align}
\begin{split}
e_x^\ast=\frac{4(\hat{\sigma}^2+\lambda)}{(x-\hat{u})^2+2\hat{\sigma}^2+2\lambda},
\label{eq:min_energy}
\end{split}
\end{align}
where $\hat{\mu}=\frac{1}{H \times W}\sum_{i=1}^{H\times W}x_i$, $\hat{\sigma}^2=\frac{1}{H \times W}\sum_{i=1}^{H\times W}(x_i-\hat{\mu})^2$, and $\lambda$ is a hyper parameter.
Each neuron within a channel shares the statistics $\mu$ and $\sigma$, which hence significantly lowers computation costs.
Given that research in neuroscience demonstrates an inverse relationship between the energy of $e_x^*$ and the significance of each neuron $x$ \cite{webb2005early}, the refinement process of a feature map can be written as,
 \vspace{-1ex}
\begin{align}
\begin{split}
\hat{\vct{x}}=\sigma(\frac{1}{\mathbf{E}})\otimes\vct{x},
\label{eq:energy_dot_prod}
\end{split}
\end{align}
where $\mathbf{E}$ groups all energy values of $e^\ast_x$, with $\sigma(\cdot)$ denoting the sigmoid function. In this study, we inserted a SimAM after the first convolution layer in each residual block of the base model.

\subsection{binary classification loss}
In this section, the fundamental cross-entropy loss with softmax and angular margin-based losses are discussed, and the weighted additive angular margin loss is proposed for our binary classification. During training, each mini-batch contains $N$ utterances from either spoofed or genuine speech, whose feature embedding vectors are $\vct{x}_i \in \mathbb{R}^D$, with the corresponding spoofing identity labels being $y_i$, where $1\le i\le N$ and $y \in \{0,1\}$ (i.e., 0 denotes spoofed speech and 1 represents the genuine).
\subsubsection{Revisiting CE-Softmax loss}
\label{sec:WCE}
The Softmax loss is comprised of a softmax function integrated with a multi-class cross-entropy loss, which is formulated as,
\vspace{-1ex}
\begin{align}
\begin{split}
&L_S=-\frac{1}{N}\sum_{i=1}^Nw_{y_i}log\frac{e^{\mathbf{W}_{y_i}^T\vct{x}_i}}{e^{\mathbf{W}_{y_i}^T\vct{x}_i}+e^{\mathbf{W}_{1-y_i}^T\vct{x}_i}}\\
&\quad=\frac{1}{N}\sum_{i=1}^Nlog(1+e^{(\mathbf{W}_{1-y_i}-\mathbf{W}_{y_i})^T\vct{x}_i}),\label{eq:softmax}
\end{split}
\end{align}
where $\mathbf{W}$ represents the weight vector of the last layer of the encoder trunk, and $\mathbf{W}_0$, $\mathbf{W}_1\in \mathbb{R}^D$ are the weight vectors of the spoofed class and genuine class, respectively. $w_{y_i}$ is the weight of the $i$-th sample with label $y_i$. This loss function merely computes penalties for classification error and does not explicitly encourage intra-class compactness or inter-class separation.
\subsubsection{Angular Margin-based loss}
The softmax loss can be reformulated so that the posterior probability only hinges on the cosine value of the angle between the weights and input vectors. With normalized unit vectors of $\widehat{\mathbf{W}}$ and $\hat{\mathbf{\vct{x}}}$, the loss function termed as Normalized Softmax Loss (NSL), is written as,
\begin{align}
\begin{split}
&L_N=-\frac{1}{N}\sum_{i=1}^Nlog\\
&\frac{e^{\mid\mathbf{\widehat{W}}_{y_i}^T\mid\mid\hat{\vct{x}}_i\mid cos(\theta_{y_i,i})}}{e^{\mid\mathbf{\widehat{W}}_{y_i}^T\mid\mid\hat{\vct{x}}_i\mid cos(\theta_{y_i,i})}+e^{\mid\widehat{\mathbf{W}}_{1-y_i}^T\mid\mid\hat{\vct{x}}_i\mid cos(\theta_{1-y_i,i})}}\\
&\quad=\frac{1}{N}\sum_{i=1}^Nlog(1+e^{(cos(\theta_{1-y_i,i})-cos(\theta_{y_i,i}))}),
\end{split}
\end{align}
where $cos(\theta_{y_i,i})$ denotes the dot product of normalized vector $\widehat{\mathbf{W}}$ ($\mid\widehat{\mathbf{W}}\mid=1$) and $\hat{\vct{x}}_i$ ($\mid\hat{\vct{x}}_i\mid=1$). Next, $\vct{x}_o=W_{y_i}^T\vct{x}_i+b_{y_i}$ describes the final linear transformation, where $x_i \in \mathbb{R}^D$ is the penultimate linear layer's output (i.e., $D$-dim embedding) of the $i$-th sample with label $y_i$ and $x_o \in \mathbb{R}^2$ is the last linear layer's output. Finally, $W_{y_i}\in\mathbf{R}^D$ denotes the $y_i$-th column of the weight $W\in\mathbf{R}^{D\times 2}$ and $b_{y_i}$ is the bias term. This bias term $b_{y_i}$ is set to 0 here, therefore, the linear transformation is reformulated as $W_{y_i}^T\vct{x}_i=\mid W_{y_i}\mid \mid \vct{x}_i \mid cos \theta_{y_i,i}$, where $\theta_{y_i,i}$ is the angle between the weights and the input feature.
Likewise, this loss function has the same issue as the Softmax loss in that it only computes penalties based on classification error. This results in embeddings which are learned by the NSL as not being sufficiently discriminative. Modifications are proposed here to mitigate this issue, where an additive margin is introduced with the AM-Softmax to make the embedding space of the two classes close to their weights $\mathbf{W}_0-\mathbf{W}_1$ and $\mathbf{W}_1-\mathbf{W}_0$. The formula for the AM-Softmax (CosFace) can now be written as,
\begin{align}
\begin{split}
&L_C=-\frac{1}{N}\sum_{i=1}^Nlog\\
&\frac{e^{s(\mid\mathbf{\widehat{W}}_{y_i}^T\mid\mid\hat{\vct{x}}_i\mid cos(\theta_{y_i,i})-m)}}{e^{s(\mid\mathbf{\widehat{W}}_{y_i}^T\mid\mid\hat{\vct{x}}_i\mid cos(\theta_{y_i,i})-m)}+e^{s\mid\widehat{\mathbf{W}}_{1-y_i}^T\mid\mid\hat{\vct{x}}_i\mid cos(\theta_{1-y_i,i})}}\\
&\quad=\frac{1}{N}\sum_{i=1}^Nlog(1+e^{s(m-cos(\theta_{y_i,i})+cos(\theta_{1-y_i,i}))}),
\end{split}
\end{align}
where $s$ denotes a hyper-parameter that rescales up the gradient instead of the numerical values becoming too small within the training phase, which helps to expedite optimization. Feature maps are rescaled by $s$, where they are accordingly projected onto a hypersphere with radius $s$. 

Furthermore, an additive angular margin penalty $m$ between $\mathbf{W}_{y_i}$ and $\vct{x}_i$ is also incorporated into the equation in order to simultaneously enhance the intra-class compactness and inter-class separability, termed as the AAM-Softmax (ArcFace) loss \cite{deng2019arcface}, formulated as,
\begin{align}
\begin{split}
&L_A=-\frac{1}{N}\sum_{i=1}^Nlog\\
&\frac{e^{s(\mid\mathbf{\widehat{W}}_{y_i}^T\mid\mid\hat{\vct{x}}_i\mid cos(\theta_{y_i,i}+m))}}{e^{s(\mid\mathbf{\widehat{W}}_{y_i}^T\mid\mid\hat{\vct{x}}_i\mid cos(\theta_{y_i,i}+m))}+e^{s\mid\widehat{\mathbf{W}}_{1-y_i}^T\mid\mid\hat{\vct{x}}_i\mid cos(\theta_{1-y_i,i})}}\\
&\quad=\frac{1}{N}\sum_{i=1}^Nlog(1+e^{s(cos(\theta_{1-y_i,i})-cos(\theta_{y_i,i}+m))}).\label{eq:aam}
\end{split}
\end{align}

\begin{figure*}[tbp]
\centering
\includegraphics[scale=0.8]{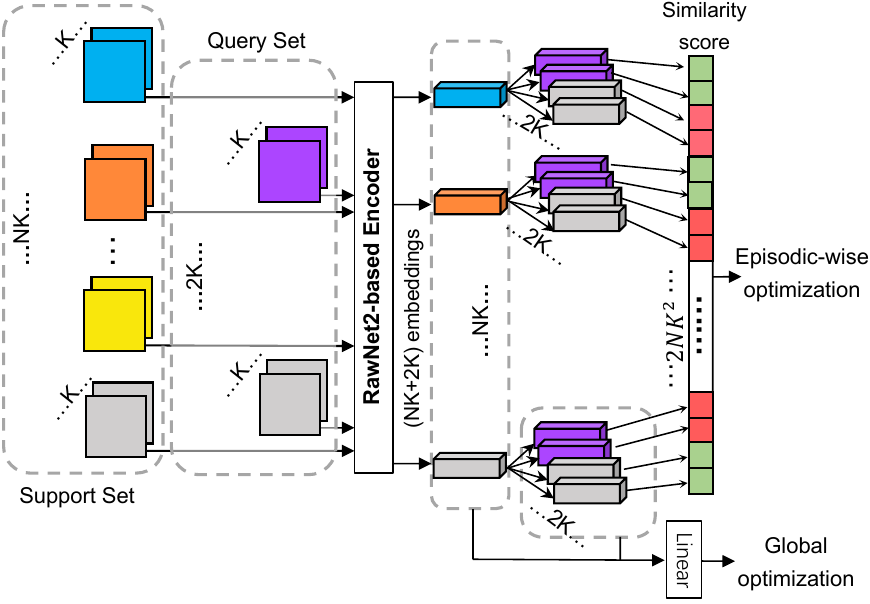}
\caption{Joint optimization scheme. All spoofing samples and embeddings are color-coded to represent different types of spoofing attacks, while genuine speech is gray. The similarity score in green denotes a match: $r_{i,j}=1$, likewise, those in red are unmatched: $r_{i,j}=0$.}
\label{fig:se2}
\vspace{-3.5ex}
\end{figure*}

\subsubsection{Weighted Additive Angular Margin loss for binary classification}
\label{sec:waam}
Since the dataset employed in this study is unbalanced (e.g., genuine versus spoofed), different classes are expected to possess individual weights for loss calculation. Here, $w_{y_i}$ denotes a manual rescaling weight assigned to class $y_i$. By adding this weight factor into the equation, a benefit is possible for the scenario when the training set is unbalanced (e.g., more spoofing samples are included versus genuine samples). Inspired by earlier research \cite{zhang2021one}, the different additive angular margin penalty $m_{y_i}$ can be injected into the corresponding target angle, which prevents the model from overfitting unseen spoofing attacks to known attacks. Specifically, there exists a distribution mismatch for spoofing attacks in the training and evaluation partition. Two different margins are therefore assigned to the bona-fide speech and spoofing attacks, which encourages better compactness for bona-fide samples, and at the same time greater isolation of the spoofing attacks. The AAM-softmax (see Eq. \ref{eq:aam}) is hence reformulated as,
\begin{align}
\begin{split}
&L_{W}=-\frac{1}{N}\sum_{i=1}^{N}log\frac{w_{y_i}\,e^{s(cos(\theta_{y_i,i}+m_{y_i}))}}{e^{s(cos(\theta_{y_i,i}+m_{y_i}))}+e^{s\, cos\theta_{(1-y_i,i)}}}\\
&=\frac{1}{N}\sum_{i=1}^Nlog\,w_{y_i}(1+e^{s(cos(\theta_{1-y_i,i})-cos(\theta_{y_i,i}+m_{y_i}))}).\label{eq:waam}
\end{split}
\end{align}

\subsection{Meta-learning episodic optimization}
\label{subsec:meta}

Meta-learning is focused on developing a task-oriented model to enhance the learning ability by conducting optimization within each subtask (i.e., an episode or a mini-batch), instead of overall engagement for a given problem. A meta-subtask is composed of a support set and a query set. Examples in the support set are used for learning how to directly solve a subtask, while the query set is used for subtask performance assessment. At each step in meta-learning, model parameters are updated based on a randomly selected subtask. Since the network is presented with various tasks at each iteration, this enforces learning to distinguish inhomogeneous examples in general, rather than a specific subset of examples. In realistic settings of the spoofing detection, training data would contain N different types of spoofing attacks manipulated by various spoofing techniques (e.g. A01-A06 in the ASVspoof 2019 logical access (LA) dataset \cite{wang2020asvspoof,todisco2019asvspoof}), but the unseen attacks could still occur in the evaluation phase. To simulate this situation during training, we first randomly select K spoofing examples $\vct{x}^s$ from each spoofing type respectively, along with 2K bona-fide examples $\vct{x}^b$. Next, one spoofing type is randomly included in the query set while keeping all other types in the support set within each subtask. Here, 2K bona-fide examples are equally distributed between the query and support set. As a result, we obtain the following support set $\mathcal{S}=\{\vct{x}^s_i\}_{i=1}^{(N-1)\times K}\cup\{\vct{x}_i^b\}_{i=1}^{K}$ and query set $\mathcal{Q}=\{\vct{x}^s_j\}_{j=1}^{K}\cup\{\vct{x}_j^b\}_{j=1}^{K}$. Given this formulation of support and query pairs in each episode, with a finite number of spoofing types of spoofing attacks enrolled into the model, the spoofing attack types in the query set can now vary in each subtask.

To compare samples in the support set and query set, we use the relation network \cite{sung2018learning}, which parameterizes the non-linear similarity metric using a neural network. Specifically, the relation network simultaneously models the feature representation and metric over a set of subtasks in order to generalize to unseen spoofing attacks. Given the input sample and its corresponding label in terms of $(\vct{x},y)$, samples from the support set $\mathcal{S}$ and query set $\mathcal{Q}$ are fed through the encoder $f_\theta$ (see Sec. \ref{sec:encoder}). Next, an embedding $f_\theta(\vct{x}_i)$ from the support set and an embedding $f_\theta(\vct{x}_j)$ from the query set are concatenated to formulate one pair. Considering the number of samples in $\mathcal{S}$ ($\vert \mathcal{S}\vert=NK$) and $\mathcal{Q}$ ($\vert\mathcal{Q}\vert=2K$), each subtask/mini-batch is comprised of $2NK^2$ permutations as a set $\mathcal{P}$ of pairs for metric-based meta-learning. Finally, each pair is processed by the relation module $f_\phi$, which yields a scalar relation output score representing the similarity between the feature representation pair,

\begin{align}
r_{i,j}=f_\phi([f_\theta(\vct{x}_i),f_\theta(\vct{x}_j)]),
\end{align}
where $[.,.]$ denotes the concatenation operation, the network $f_\phi$ treats the relation score as a similarity measure\cite{sung2018learning}, therefore $r_{i,j}$ is defined as,

\begin{align}
\label{eq6}
r_{i,j}=\left\{
\begin{aligned}
1, \ \ if\ y_i=y_j, \\
0, \ \ otherwise.
\end{aligned}
\right.
\end{align}
The network $f_\theta$ and $f_\phi$ are jointly optimized using mean square error (MSE) objective as in \cite{sung2018learning}, where the relation network output is treated as the output of a linear regression model. The MSE loss for meta-learning here is written as,
\begin{align}
L_{M}=\frac{1}{2NK^2}\sum_{i=1}^{NK}\sum_{j=1}^{2K}(r_{i,j}-1(y_i==y_j))^2.\label{eq:mse}
\end{align}

Additionally, we enforce the model to classify samples in both the support and query sets against the entire set of classes in the training set. The entire meta-learning scheme with global classification is depicted in Fig. \ref{fig:se2}. A hyper-parameter $\lambda$ balances the weighted AAM loss (Eq. \ref{eq:waam}) and the MSE loss, where the fusion loss is hereby written as,
\begin{align}
L_{F}=L_{W}+\lambda L_{M}.\label{eq:fusion}
\end{align}

\subsection{Disentangled adversarial training}
\label{sec:disentangle}
\subsubsection{Adversarial examples}

Next, adversarial examples can be obtained by adding imperceptible but malicious perturbations to the original training data, which can compromise the accuracy of a well-trained neural network \cite{goodfellow2014explaining}. Adversarial examples are commonly treated as a threat to neural networks. Here, we leverage both original training data and corresponding adversarial examples to train networks for enhanced system performance. Consider the default adversary generation method, the Fast Gradient Sign Method (FGSM), which has random perturbation and has been used for maximizing the inner part of the saddle point formulation \cite{tramer2017space}. A more powerful multi-step attacker based on the projected gradient descent (PGD) (see Eq. \ref{eq:pgd}) is adopted here to produce adversaries on the fly \cite{madry2017towards}. Given an input training sample $\vct{x} \in \mathbb{D}$ with a corresponding ground-truth label $y$, adversary generation is conducted in an iterative manner as follows,

\begin{align}
\begin{split}
\vct{x}_{t}^{adv}=\Pi_{\vct{x} + \mathbb{S}}(\vct{x}_{t-1}^{adv}+\mathop{\alpha} \text{sgn}( \nabla_{\vct{x}}\mathrm{L}(\theta,\vct{x},y))),
\label{eq:pgd}
\end{split}
\end{align}
where $\Pi$ denotes a projection operator, $\mathbb{S}$ represents the allowed perturbation size that formalizes the manipulative power of the adversary, $\alpha$ is the step size, $\mathrm{L}(\cdot,\cdot,\cdot)$ stands for the loss function, and $\theta$ indicates the model parameters. Eq. \ref{eq:pgd} then illustrates one step of a multi-step attacker to generate adversaries. 

The adversarial training framework proposed in \cite{madry2017towards} only used maliciously perturbed samples to train networks. Here, the robust optimization objective illustrates a saddle point problem composed of an inner maximization problem and an outer minimization problem written as,
\begin{align}
\begin{split}
\mathop{\arg\min}_{\theta}\mathbb{E}_{(\vct{x},y)\sim\mathbb{D}}(\mathop{\max}_{\delta\in\mathbb{S}}\mathrm{L}(\theta,\vct{x}+\delta,y)).
\label{eq:pgd_opt}
\end{split}
\end{align}
For each training data sample $\vct{x} \in \mathbb{D}$, a set of allowed perturbations $\delta \in \mathbb{S}$ are introduced to formalize adversaries. Such a training framework has merits as described in \cite{freelunch,yin2019fourier,zhang2019interpreting}, but cannot generalize well to original training data \cite{madry2017towards,xie2019feature}.

To encourage full exploitation of the complementarity nature between original training data and corresponding adversarial examples, adversarial examples are treated as augmented data, and incorporated with the original data for model training. The learning objective is formulated as,
\begin{align}
\begin{split}
&\mathop{\arg\min}_{\theta,\phi}\mathbb{E}_{(\vct{x},y)\sim\mathbb{D}}(\mathrm{L}_{F}(\theta,\phi,\vct{x},y))\\
&+\mathop{\arg\min}_{\theta}\mathbb{E}_{(\vct{x},y)\sim\mathbb{D}}(\mathop{\max}_{\delta\in\mathbb{S}}\mathrm{L}_W(\theta,\vct{x}+\delta,y),
\label{eq:learning_obj}
\end{split}
\end{align}
where $\mathrm{L}_{F}$ and $\mathrm{L}_W$ is referred to as Eq. \ref{eq:fusion} and Eq. \ref{eq:waam}, respectively. 

\subsubsection{Disentangling via an auxiliary BN}
\label{subsec:DAT}

\begin{figure}[h]
\centering
\includegraphics[scale=0.4]{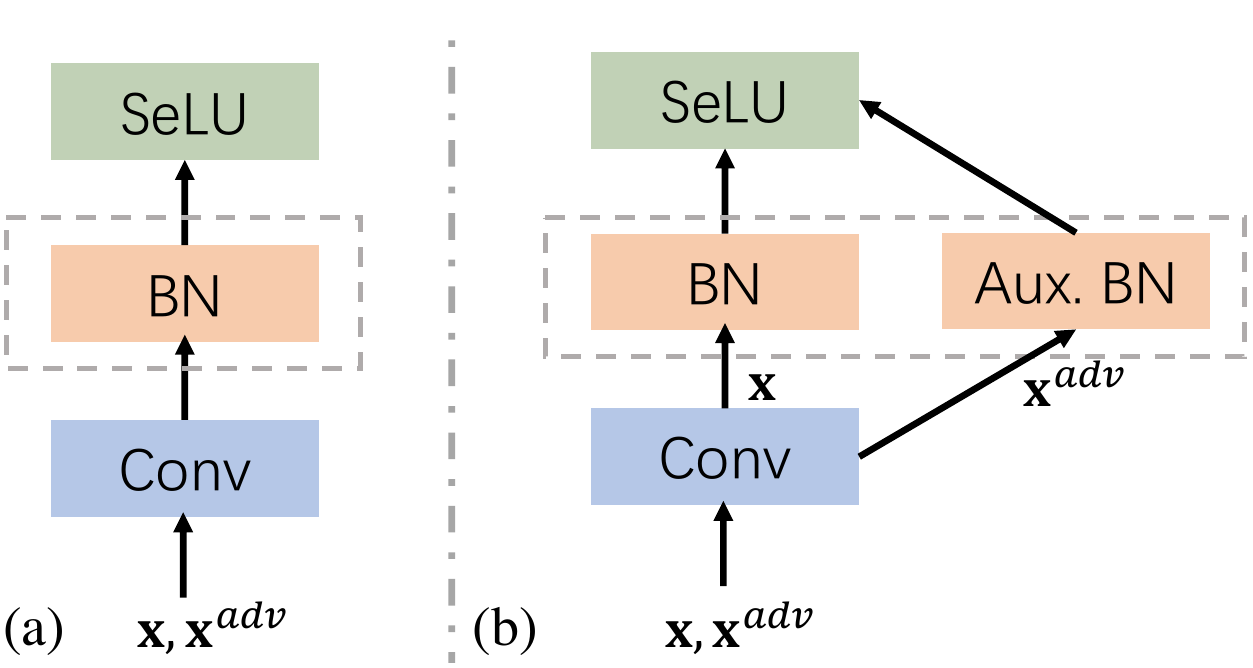}
\caption{Alternate data flow options between architectures with conventional BN (a) and with auxiliary BN (b)}
\label{fig:aux_bn}
\vspace{-3.5ex}
\end{figure}

Earlier studies on adversarial attacks have demonstrated that training using adversarial examples can cause label leaking (i.e., the neural network overfits to the specific adversary distribution), which leads to compromised model performance \cite{kurakin2016adversarial} \cite{goodfellow2014explaining}.
Under the assumption that adversarial examples and original data come from different underlying distributions, Xie et al. proposed disentangled training via an auxiliary batch norm (BN) to decouple the batch statistics between original and adversarial data in the normalization layers during model training \cite{xie2020adversarial}. This approach would allow for better exploitation of the regularization power of adversarial data. For the original mini-batch training data and corresponding adversarial data at each training step, we hereby utilize two BNs (i.e., one main BN and one auxiliary BN) for specific data partitions while the remaining model parameters are jointly tuned. Corresponding data flows in different architectures (i.e., with conventional BN and with Auxiliary BN) are illustrated in Fig. \ref{fig:aux_bn}. 
At the evaluation phase, we maintain only the main BN for data distribution normalization while bypassing the auxiliary one.
\vspace{-1ex}{}

\subsubsection{Adversarial training scheme}
Compared to adversarial training \cite{goodfellow2014explaining,kurakin2016adversarial}, disentangled learning can fully exploit the complementarity nature between original training data and corresponding adversarial examples. Adversarial examples are generated during model training. In each training iteration, we treat the original data mini-batch as adversarial examples at the initial step (i.e., $t=0$). Multiple steps for attacking are performed using the auxiliary BNs. We then derive the adversaries for the current mini-batch. The objective of incorporating adversarial examples into training is to improve the model generalization ability to unseen spoofing attacks. We thereby substitute adversaries corresponding to the bona-fide label (i.e., $y^{adv}=1$) with original training samples to maintain an identical data distribution of genuine samples in the adversarial mini-batch. Subsequently, we submit the original mini-batch and adversarial mini-batch to the same network, while calculating the loss via main BNs and auxiliary BNs for specific mini-batches, respectively. Finally, we minimize the total loss for the network parameter updates (Eq. \ref{eq:learning_obj}). We present the complete training scheme with adversarial examples in Algorithm \ref{alg:at}.
\begin{algorithm}[h]
\captionsetup{font={footnotesize}}
  	\caption{ Adversarial training scheme}
  	\label{alg:at}
  	\begin{algorithmic}[1]
  	\begin{footnotesize}
  		\Require Original training data with labels $\lbrace\vct{x}^{org},\vct{y}^{org}\rbrace\sim\mathbb{R}$
  		\Ensure	Encoder parameter $\theta$; relation module parameter $\phi$
  		
  		\State Given $S$ training iterations, $T$ attacking steps, and batch size $N$
  		\For{$s$ in $1:S$}
  		\State sample mini-batch $\lbrace\vct{x}_i^{org},y_i^{org}\rbrace_{i=1}^m$ 
  		
  		\State Suppose $\lbrace\vct{x}_{i,0}^{adv},y_{i,0}^{adv}\rbrace_{i=1}^m=\lbrace\vct{x}_i^{org},y_i^{org}\rbrace_{i=1}^m$ 
  		\For{$t$ in $1:T$}
  		\State Generate adversarial examples $\lbrace\vct{x}_{i,t}^{adv},y_{i,t}^{adv}\rbrace_{i=1}^m\sim\mathbb{R}\cup\mathbb{S}$ at current time step $t$ using auxiliary BNs w/ Eq. \ref{eq:pgd}
  		\EndFor
  		\For{$n$ in $1:N$}
  		\If{$y_n^{adv} = 1$}\Comment{label 1 represents the bona-fide}
  		\State $\vct{x}^{adv}_n = \vct{x}^{org}_n$ \Comment{maintain original data for genuine samples}
  		\EndIf
  		\EndFor
  		\State \textbf{Compute} $\mathrm{L}_F(\vct{x}^{org},\vct{y})$ w/ Eq. \ref{eq:fusion}
        \State \textbf{Compute} $\mathrm{L}_W(\vct{x}^{adv},\vct{y})$ w/ Eq. \ref{eq:waam}
  		\State \textbf{Update} $\theta$ and $\phi$ w/ Eq. \ref{eq:learning_obj}
  		\EndFor
  	\end{footnotesize}
  	\end{algorithmic}
\end{algorithm}

\section{Experiment}
\label{experiment}

\subsection{Dataset and evaluation Metrics}
The ASVspoof 2019 corpus on the Logical Access (LA) track \cite{todisco2019asvspoof,wang2020asvspoof} is adopted in this work to train and test models. The corpus consists of three partitions, namely, training, development, and evaluation subsets, with each subset containing genuine and spoofed samples. Different spoofing methods (i.e., voice conversion and speech synthesis) are employed to create spoofing attacks \cite{yamagishi2019asvspoof}. The evaluation partition features 13 different attacking genres (A07-A19); the training and development subsets contain 6 different spoofing attacks (A01-A06). Model selection and gauging emergence of the over-fitting are dependent on the development subset. Given the 13 algorithms used for generating evaluation data, 2 algorithms are also used in training and development subsets, while the other 11 algorithms are unseen/uninvolved for train and development data. Bona-fide samples are collected from 107 speakers. The number of audio samples in each subset are 25,380, 24,986, 71933 for training, development, and evaluation, respectively. The durations of each speech sample ranges from 1-2 sec, with all audio samples in each subset stored in flac format. 

We adopt the equal error rate (EER) and the minimum normalized tandem detection cost function (min t-DCF) \cite{kinnunen2018t,kinnunen2020tandem} as the metrics for assessing system performance. Wang et al. \cite{wang2021comparative} found that spoofing detection systems initialized with different random seeds can deliver different results by a substantial margin. As such, all results reported here are the best results from three runs with different random seeds.

\subsection{Implementation Details}

\label{sec:model_conf}
The currently proposed spoofing detection system is implemented using Pytorch toolkit. Each input segment is approximately 4 sec in duration, and processed by a RawNet2-based encoder \cite{Jung2020ImprovedRW}. The RawNet2-based encoder consists of a sinc-convolution layer \cite{ravanelli2018speaker} and six stacked residual blocks with pre-activation \cite{he2016identity}. The sinc-convolution layer is initialized with a bank of 70 mel-scaled filters. Each residual block is stacked with a batch normalization layer \cite{ioffe2015batch}, a scaled exponential linear unit (SeLU) activation \cite{klambauer2017self}, a 2D convolution layer, and a max pooling layer. The first two residual blocks are equipped with 32 filters, while the remaining four have 64 filters. After the encoder, there is an adaptive average pooling layer to aggregate frequency-wise information. Next, a gated recurrent unit (GRU) with 64 hidden units is used to aggregate sequential features within the temporal domain. The intermediate features are then processed using a fully connected layer with 64 units. The 64-dim embeddings extracted at the final layer are subsequently used for calculating similarity scores and estimating classification loss. The relation network has two fully connected layers with 64 units each. Additionally, we employ Projected Gradient Descent (PGD) \cite{madry2017towards} under an $\mathrm{L}_{\infty}$ norm as the default attacker for crafting adversarial examples on-the-fly. The perturbation size $\delta$ (see Eq. \ref{eq:pgd_opt}) is set to 0.002. The number of attacking iterations is set to 12. The attack step size (see Eq. \ref{eq:pgd}) is fixed to $\alpha=0.0001$, and the balance hyper-parameter $\lambda$ in Eq. \ref{eq:fusion} is set to 0.8.

We conducted extensive experiments using multiple setup combinations with loss functions, attention modules, and disentangled training with adversarial examples. The baseline system employs the RawNet2-based encoder to learn the spoofing identity, which minimizes a cross-entropy loss w.r.t the network parameters for gradient updates. The ASVspoof 2019 corpus is data-unbalanced with a 1:9 ratio of genuine samples to spoofing samples, thereby assigning specific weights to genuine and spoofing classes with 0.1 and 0.9, respectively. Likewise, category-wise weights $w_{y_i}$ in Eq. \ref{eq:waam} are designated in the same way. Also, there are two hyper-parameters in Eq. \ref{eq:waam}, where the scale $s$ is fixed to 32, while the margin $m_0,m_1$ are set to 0.2 and 0.9, respectively. The batch size in each experiment is set to 16. During the meta-learning sampling phase in one episode/mini-batch, we randomly select 2 ($K=2$, see Sec. \ref{subsec:meta}) samples from each attacking type (A01-A06) and 4 samples from genuine samples (4 genuine samples are equally split into the support and query sets). With regard to our model optimization strategy, we utilize the Adam optimizer \cite{kingma2014adam} with a learning rate of 0.0001 using a cosine annealing learning rate decay. The model in each experiment was trained for 100 epochs. For the SimAM attention module (see Sec. \ref{sec:simam}), the hyper-parameter $\lambda$ in Eq. \ref{eq:min_energy} is set to 0.0001.

\section{Result and Analysis}
\label{res}
To thoroughly evaluate our proposed methods, we assess the feature enhancement effectiveness for different attention modules, then search for a rational perturbation size to craft adversarial examples, and present an ablation study on loss functions, and a comparison of our results to the state-of-the-art systems in this section.
\subsection{attention module selection}
The RawNet2-based encoder model learns high-level representations for spoofing identities, while there are several attention modules that can be leveraged to refine the intermediate feature maps. As noted in Sec. \ref{sec:model_conf}, the baseline system uses a RawNet2-based encoder, which is trained with a cross-entropy (CE) loss (see Sec. \ref{sec:WCE}). We compare spoofing detection system performances derived from the encoder (see Sec. \ref{sec:encoder}) equipped with different attention modules. The results are presented in terms of min t-DCF and EER in Tab. \ref{table:attention}.

As results shown in Tab. \ref{table:attention}, the baseline system achieves acceptable results in terms of min t-DCF and pooled EER, which is probably owing to the interpretation of single-channel 2-dim feature map generated by the sin-convolution layer, thereby enhancing the feature representation ability. Each attention module improves system performance to varying degrees, which means they contribute to refining the intermediate feature maps. The system with CBAM yields a lower EER, but the SE module outperforms CBAM in terms of min t-DCF. The system with CBAM delivers a better spoofing detection performance than that with SE while resulting in a higher expected detection cost. Compared to the previous two attention modules, the SimAM encourages learning the informative feature maps along with superior system performance.

\begin{table}
\centering
  \renewcommand{\arraystretch}{1.1}
  \caption{The effectiveness for different attention modules}
  \vspace{-2ex}
    \begin{tabular}{ccc}
    \hline
System & min t-DCF & EER (\%) \\
    \hline
    \hline
    RawNet2 + CE (baseline) & 0.0566 & 1.67 \\
    \hline
RawNet2 w/ SE before BN & 0.0492 & 1.64 \\
     \textbf{RawNet2 w/ SE after BN} & 0.0412 & 1.62 \\
RawNet2 w/ CBAM before BN & 0.0514 & 1.55 \\
     \textbf{RawNet2 w/ CBAM after BN} & 0.0456 & 1.52 \\
     \textbf{RawNet2 w/ SimAM before BN} & \textbf{0.0406} & \textbf{1.41} \\
    RawNet2 w/ SimAM after BN & 0.0458 & 1.43 \\
    
    \hline
    \end{tabular}%
    \label{table:attention}
    \vspace{-2ex}
\end{table} 
We found that different insertion positions of each attention module can result in varied spoofing detection performances. Inserting the attention module right after/before the BN in the residual block improves distinctive feature learning. In each residual block, either the SE or CBAM encourages feature refinement more significantly, while inserted right after BN, and SimAM conducts the more effective feature enhancement while inserted after the first convolutional layer and before the BN. The best result in this section is delivered with the RawNet2-based encoder equipped with SimAM, while the module is inserted before the BN in each residual block. The EER is reduced to 1.41\% with $+$ 15.57\% relative reduction compared to the result of the baseline system, and min t-DCF improves to 0.0406 with a $+$ 28.27\% relative reduction.

\subsection{Search for optimal perturbation size}

During adversaries generation, a set of allowed perturbations $\delta \in \mathbb{S}$ (see Eq. \ref{eq:pgd_opt}) formalize the manipulative power of adversarial examples. We investigate multiple orders of magnitude of perturbation size on the effectiveness of enhancing model robustness/accuracy. 
\begin{table}[ht]
\centering
  \renewcommand{\arraystretch}{1.1}
  \caption{The effectiveness for different perturbation size}
  \vspace{-2ex}
    \begin{tabular}{ccc}
    \hline
System & min t-DCF & EER (\%)\\
    \hline
    \hline
    RawNet2 + CE (baseline) & 0.0566 & 1.67 \\
    \hline
RawNet2 + CE + Adv. ($\delta=0.1$) & 0.0649&2.06\\  
RawNet2 + CE + Adv. ($\delta=0.01$) & 0.0564&1.69\\
RawNet2 + CE + Adv. ($\delta=0.001$) & 0.0372&1.33\\
RawNet2 + CE + Adv. ($\delta=$\textbf{0.002}) & \textbf{0.0356} & \textbf{1.142}\\
RawNet2 + CE + Adv. ($\delta=0.003$) & 0.376&1.43\\
RawNet2 + CE + Adv. ($\delta=0.004$) & 0.468&1.55\\
RawNet2 + CE + Adv. ($\delta=0.0001$) & 0.0514& 1.63\\
    
    \hline
    \end{tabular}%
    \label{table:perturb}
    %\vspace{-2ex}
\end{table} 
\begin{table*}[ht]
\centering
  \renewcommand{\arraystretch}{1.1}
  \caption{Breakdown EER performance of 13 attacks in the ASVspoof 2019 LA evaluation partition, pooled min t-DCF, and pooled EER,The best performance for each column is marked in boldface.}
  \vspace{-2ex}
  \setlength{\tabcolsep}{1.8mm}{
    \begin{tabular}{c||ccccccccccccc||cc}
    \hline
    System & A07 & A08 & A09 & A10 & A11 & A12 & A13 & A14 & A15 & A16 & A17 & A18 & A19 & min t-DCF & EER (\%)\\
    \hline 
    Baseline & 1.79 & 0.69 & 0.06 & 2.42 & 1.28 & 2.73 & 0.38 & 0.75 & 1.12 & 0.61 & 1.81 & 3.52 & 1.33 & 0.0566 & 1.67 \\
     RawNet2 w/ SimAM + CE &\textbf{0.57}&0.41&0.04&0.91&0.34&1.96&0.12&0.34&\textbf{0.42}&0.67&1.85&3.58&1.23& 0.0406 & 1.41\\
    \hline
    Replace CE w/ AAM  &0.92&0.27&0.02&1.27&0.33&1.83&0.15&0.26&0.69&0.68&2.09&2.67&1.08& 0.0399 & 1.36 \\
    Replace CE w/ WAAM  &1.30&\textbf{0.14}&\textbf{0.00}&1.65&\textbf{0.31}&1.70&0.19&0.14&0.96&0.65&2.34&1.77&0.91& 0.0389 & 1.29 \\
    + MSE   & 0.91 & 0.29 & 0.02 & 1.39 & \textbf{0.31} & 1.30 & \textbf{0.08} & 0.18 & 0.61 & \textbf{0.24} & 2.11 & 2.25 & 0.97 & 0.0289 & 0.99\\
    + Adv.  & 0.63 & 0.22 & \textbf{0.00} & \textbf{0.85} & 0.35 & \textbf{0.91} & 0.34 & \textbf{0.12} & 0.89 & 0.75 & \textbf{1.79} & \textbf{1.58} & \textbf{0.63} & \textbf{0.0277} & \textbf{0.87}\\
      
    \hline
    \end{tabular}}%
    \label{table:breakdown}
    \vspace{-3ex}
\end{table*}  

As shown in Tab. \ref{table:perturb}, spoofing detection performances are compromised while training with slightly larger perturbation (i.e., $\delta=0.1/0.01$), potentially due to the fact that excessive perturbations could in fact blur the distinctive original identity pattern, causing misclassification. In contrast, a slightly small perturbation size (i.e., $\delta=0.0001$) is trivial to generate strong enough adversaries in order to improve robustness. Additionally, \cite{madry2017towards} found that increasing the capacity of the network when training using only original training data improves robustness against adversaries, and this effect is greater when considering adversaries with small perturbations. Moreover, performance on the original training samples can be degraded by the small capacity of the network, providing some form of robustness against adversaries \cite{madry2017towards}. We observe that adversarial examples with a spectrum of perturbation sizes (i.e., $\delta \in [0.001,0.004]$) are exerting varying degrees of influence on boosting system performance. This is especially the case for adversaries generated with perturbation size $\delta=0.002$, which maximize contributions to system performance improvement.

\subsection{Ablation study on loss functions}

We perform an ablation study on diverse loss functions based on the RawNet2-based architecture. An ablation study serves to understand the contribution of each component to overall system performance. The base model (see Sec. \ref{sec:encoder}) equipped with SimAM (see Sec. \ref{sec:simam}) exhibits a satisfying encoding ability to absorb distinctive information from input features. Firstly, we minimize a cross-entropy (CE) loss (see Eq. \ref{eq:softmax}) w.r.t. the network parameter for gradient updates. To encourage better binary classification, we replace the CE loss with the weighted additive angular margin (WAAM) loss (see Eq. \ref{eq:waam}). Subsequently, we incorporate the meta-learning mean square error (MSE) loss (see Eq. \ref{eq:mse}) into a fused total loss function (see Eq. \ref{eq:fusion}). Additionally, to leverage the regularization power of adversarial examples, we conduct disentangled training (see Sec. \ref{sec:disentangle}) with a mixture of original training data and corresponding spoofing adversaries under a combined learning objective (see Eq. \ref{eq:learning_obj}).

A breakdown of the results on the evaluation partition with unknown attacks is illustrated in Tab. \ref{table:breakdown}. Each proposed loss function boosts system performance with incremental improvement. The WAAM loss outperforms cross-entropy loss, specifically, the relative reduction in EER is up to $+$ 8.5\%, and $+$ 4.2\% on min t-DCF. Additionally, the parameter $w_{y_i}$ in Eq. \ref{eq:waam} is beneficial to the system performance improvement, which is designed to alleviate the impact by data imbalance. The joint optimization of the WAAM loss and meta-learning MSE loss promotes better generalization to unseen spoofing attacks, yielding a better spoofing detection result. In addition, disentangled learning further facilitates distinctive embedding learning, leading to enhanced system accuracy/robustness. Revisiting the baseline system (also see in Tab. \ref{table:attention}), the best solution outperforms the baseline system by $+$ 47.9\% relative EER reduction, and $+$ 51.1\% relative min t-DCF reduction.

\begin{figure}[h]
\centering
\includegraphics[scale=0.59]{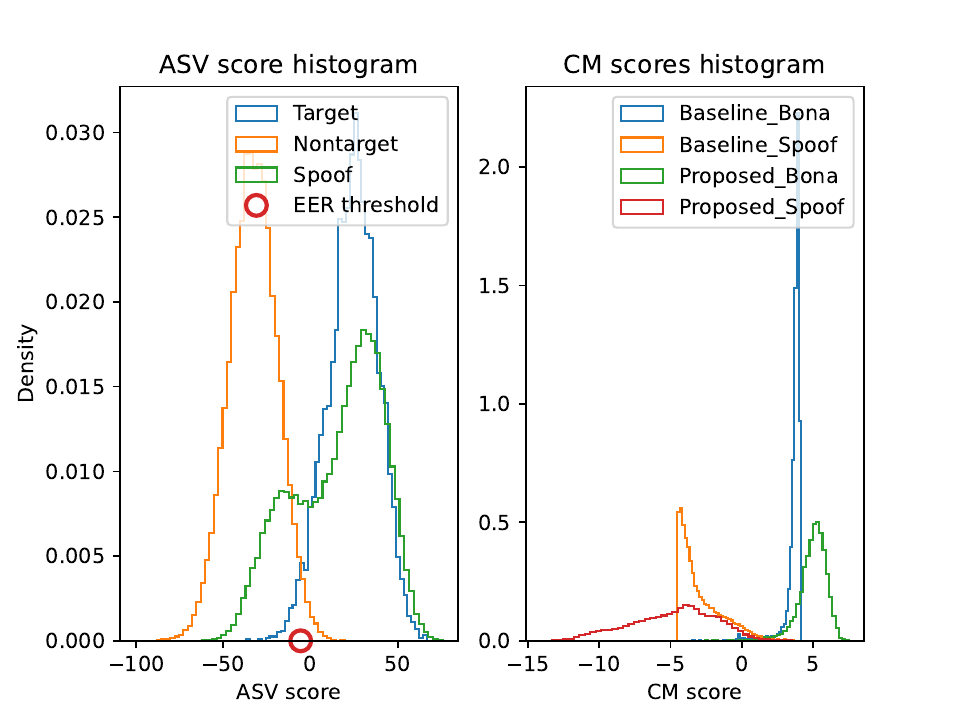}
\caption{Distribution of ASV scores and countermeasure scores.}
\label{fig:dis}

\end{figure}

\begin{figure}[h]
\centering
\includegraphics[scale=0.58]{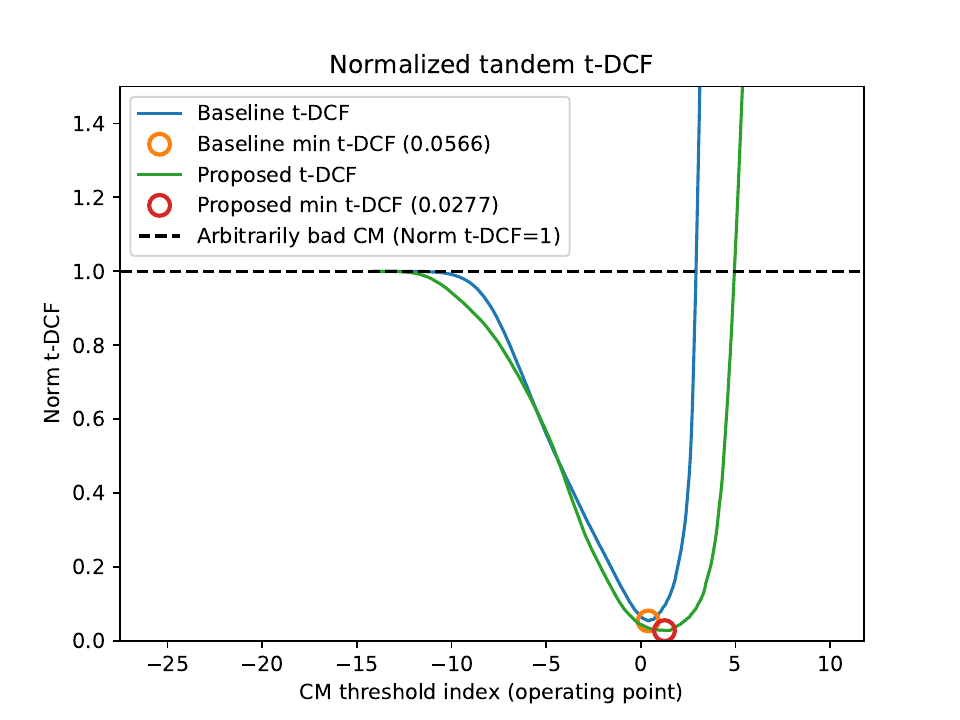}
\caption{Normalised ASV-constrained t-DCFs plot for the baseline CM system and proposed CM system on ASVspoof 2019 LA track.}
\label{fig:tdcf}

\end{figure}

\begin{figure*}[ht]
\centering
\vspace{-1.5ex}
\subfigure[Baseline]{\includegraphics[width=0.46\linewidth]{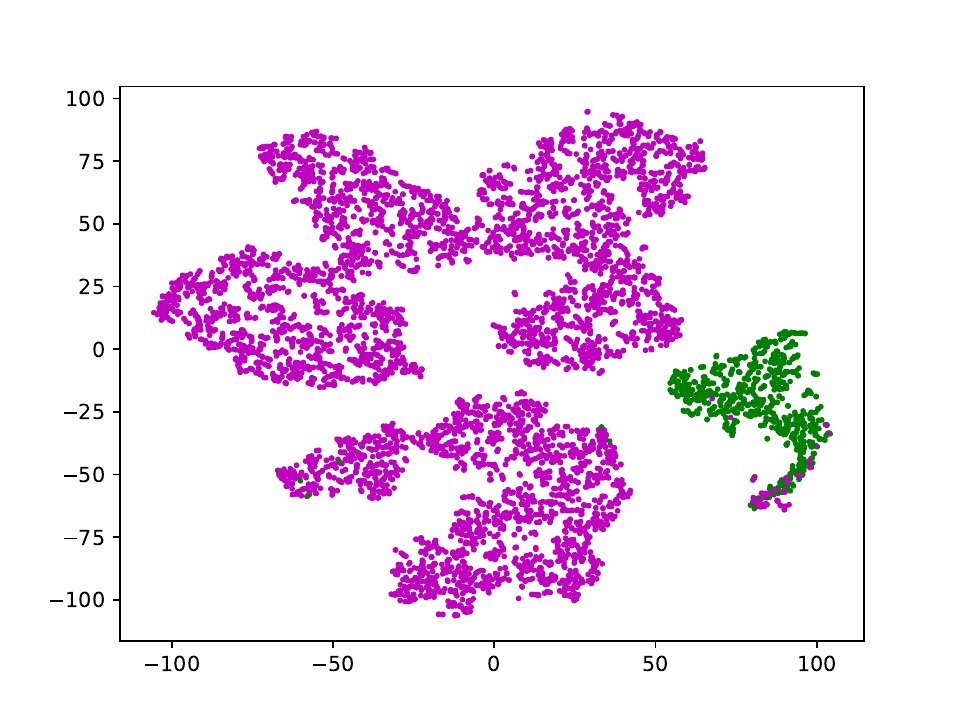}}
\subfigure[Proposed]{\includegraphics[width=0.46\linewidth]{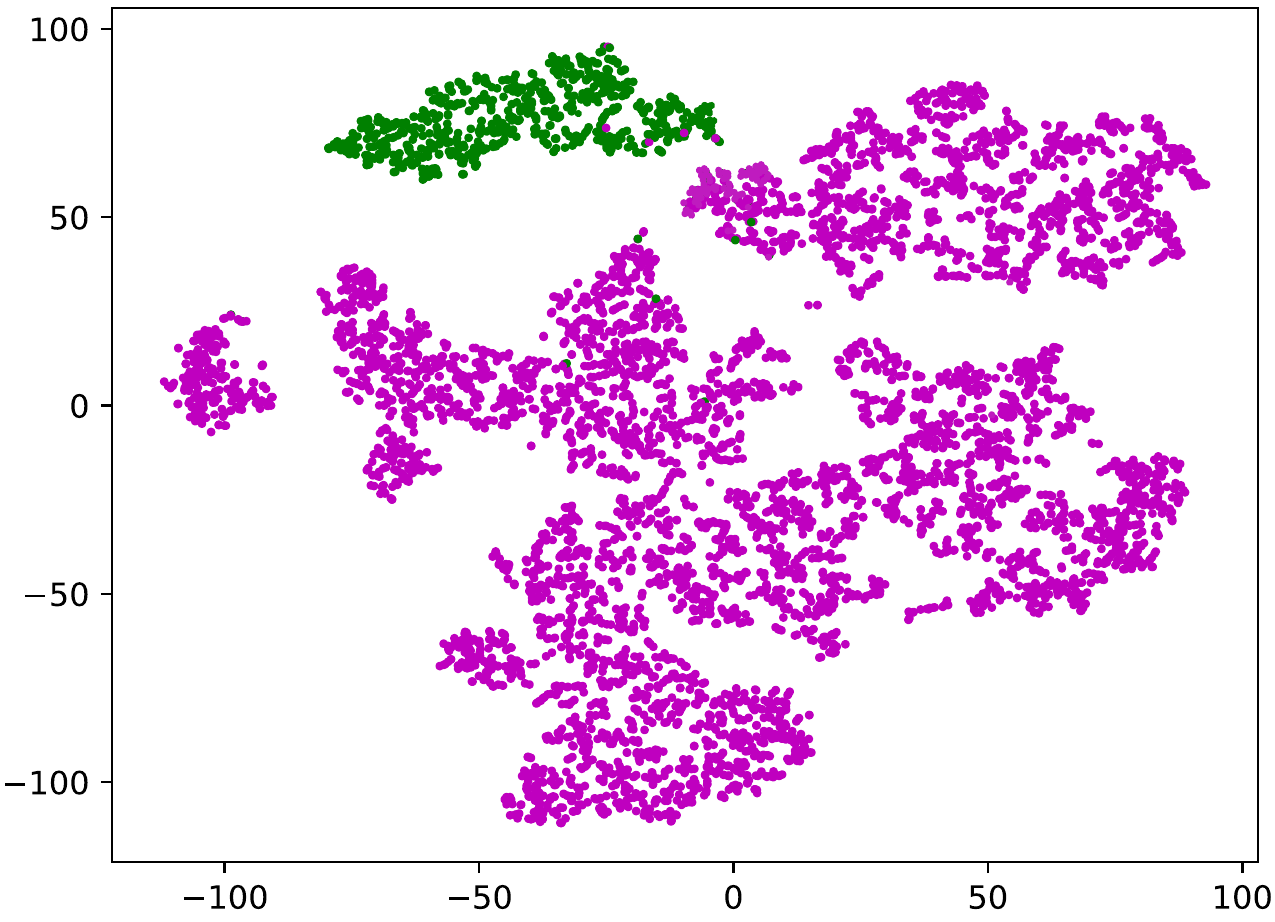}}
\captionsetup{font={footnotesize}}
\caption{t-SNE visualization of feature embedding for the baseline CM system and proposed CM system on ASVspoof 2019 LA track (genuine samples are color-coded as green, spoofed samples are present in purple).}
\label{fig:tsne} 
\vspace{-1.5ex}
\end{figure*}

\subsection{Visualization of improvement by proposed approach}

Aligning with the definition of a tandem system as a cascade of countermeasure (CM) and ASV systems in \cite{kinnunen2020tandem}, the CM acts as a gate to filter out spoofing attacks before reaching the ASV system. The tandem system can encounter three types of trials: (i) target, (ii) non-target and (iii) spoof. Only the target trials should be accepted while both non-target and spoof trials should be rejected. Fig. \ref{fig:dis} presents score density distributions, which comprises ASV scores (on the left panel) and CM scores (on the right panel) for both baseline CM system and proposed CM system. In Fig. \ref{fig:dis}, “Bona” and “Spoof” mean bona fide speech and spoofing attacks, respectively. The proposed CM system tends to yield a wider score distribution, which is consistent with the idea of better generalization by potentially encompassing scores from unseen samples. Additionally, a larger margin is observed between the genuine speech space and the spoofing space in the proposed approach, which further forces inter-class separability.

Fig. \ref{fig:tdcf} shows the t-DCFs plot which provides insight into the observation made from Fig. \ref{fig:dis}. The tandem detection cost function (t-DCF) \cite{kinnunen2018t} metric reflects the overall performance of a combined ASV and CM system. Here, the ASV-constrained normalised t-DCF curves are shown for the baseline CM system and proposed CM system, while evaluated on the ASVspoof 2019 LA track, when varying the CM threshold. The proposed CM system reached a lower minimum t-DCF than the baseline CM system, indicating a better overall spoofing detection performance. 

In order to further prove effectiveness of the proposed approach, the dimension-reduced feature embedding of the baseline CM and proposed CM are visualized in Fig. \ref{fig:tsne}. We utilize t-distributed Stochastic Neighbor Embedding (t-SNE) \cite{van2008visualizing} to visualize feature embedding for the evaluation partition on ASVspoof 2019 LA track. As shown in Fig. \ref{fig:tsne}, the proposed CM system can distinguish genuine speech and spoof attacks better than the baseline CM system with fewer misclassified spoofed samples, which indicates a better generalization ability to unseen spoofing attacks is present in the proposed CM system.

\subsection{Performance comparison against existing systems}

As illustrated in Tab. \ref{table:syscom}, a comparison of system performance between our proposed CM system and competing single state-of-the-art systems is also presented. The classical machine-learning-based method uses a common GMM back-end with LFCC as the front-end, which shows satisfying classification performance \cite{tak2020spoofing}. Comprehensive results show that our introduced simple attention module outperforms alternative approaches from previous works such as Convolutional Block Attention Module (CBAM) \cite{ma2021improved}; Squeeze-and-Excitation (SE) \cite{li2021replay,zhang122021effect}; and Dual attention module with pooling and convolution operations \cite{ma2021improved}. The WAAM loss employed in this work inherited the merits of cross-entropy loss and one-class softmax loss in \cite{zhang2021one}. Subsequently, the system trained with WAAM loss for binary classification outperforms the OC-Softmax, and AM-Softmax losses proposed in \cite{zhang2021one}. Both RawGAT-ST system \cite{tak2021end} and Raw PC-DARTS system \cite{ge2021raw} operate directly on the raw speech data, while the former system is based upon graph attention networks, the latter system suggests an interesting approach to learn the network architecture automatically. Recent works \cite{eom2022anti,lee2022representation,lin2022light} employed pre-trained Wav2Ves 2.0 as the front-end to extract speech embedding which are already learned from another task \cite{baevski2020wav2vec}, embeddings are then mapped to the latent feature via proposed networks. More recently, the Rawformer \cite{liu2023leveraging} is proposed to leverage positional-related local-global dependency for synthetic audio spoofing detection. Those approaches with pre-trained embedding extractor show good performance in low-resource and cross-dataset settings. An interesting research work has been done in \cite{ito2023spoofing}, they found that the attacker could also benefit from self-supervised learning (SSL) models (i.e., wav2vec 2.0 \cite{baevski2020wav2vec}, HuBERT \cite{hsu2021hubert}, and WavLM \cite{chen2022wavlm}), thereby eliminating most of the benefits the defender gains from them. Overall, our proposed system delivers the competitive results of all systems evaluated on the ASVspoof 2019 LA track.

\begin{table}[h]
\centering
  \renewcommand{\arraystretch}{1.1}
  \caption{Performance on the ASVspoof 2019 LA evaluation partition in terms of min t-DCF and pooled EER for state-of-the-art systems and our proposed best system.}
  \setlength{\tabcolsep}{1.4mm}{
  \vspace{-1ex}
    \begin{tabular}{cccc}
    \hline
Architecture & Front-end & min t-DCF & EER (\%)\\
    \hline
    \hline
    Ours & SincNet & 0.0277 & 0.87 \\
    Dual-Branch Network\cite{ma2023end} & CQT,LFCC & 0.021 & 0.80 \\
    WavLM+ \cite{ito2023spoofing} & HuBERT &N/A&0.23\\
    SE-Rawformer \cite{liu2023leveraging} & SincNet&0.0184& 0.59 \\ 
    SBNLCNN \cite{lin2022light} & wav2vec 2.0 & N/A & 0.258 \\
    XLSR-ASP \cite{lee2022representation} & XLSR-53 & 0.0088 & 0.31 \\
    VIB \cite{eom2022anti}& wav2vec 2.0 & 0.0107 & 0.40 \\
    
RawGAT-ST \cite{tak2021end}& SincNet & 0.0335 & 1.06 \\
SENet \cite{zhang122021effect} & FFT & 0.0368 & 1.14 \\
Raw PC-DARTS \cite{ge2021raw} & SincNet & 0.0517 & 1.77 \\
MCG-Res2Net50+CE \cite{li2021channel} & CQT & 0.0520 & 1.78 \\
ResNet18-LMCL-FM \cite{chen2020generalization} & LFB & 0.0520 & 1.81 \\
Res18-OC-Softmax \cite{zhang2021one} & LFCC & 0.0590 & 2.19 \\
SE-Res2Net50 \cite{li2021replay} & CQT & 0.0743 & 2.50 \\
LCNN-Dual attention \cite{ma2021improved} & LFCC & 0.0777 & 2.76 \\
Res18-AM-Softmax \cite{zhang2021one} & LFCC & 0.0820 & 3.26 \\
    GMM \cite{tak2020spoofing} & LFCC & 0.0904 & 3.50 \\
    LCNN-4CBAM \cite{ma2021improved} & LFCC & 0.0939 & 3.67 \\
    \hline
    \end{tabular}}
    \label{table:syscom}
\vspace{-2ex}
\end{table}

\section{Conclusion}
\label{conclusion}
This study has considered the formulation of an effective approach to improve robustness of synthetic audio spoofing detection. We employed the RawNet2-based encoder, equipped with a simple attention module for feature refinement, to strengthen the distinctive feature representation power. Subsequently, we extensively explored multiple loss functions and their fusion to calibrate an embedding space for enhanced generalization to unseen spoofing attacks. First, we put forward the weighted additive angular margin loss, which served to alleviate any data imbalance and refine the embedding distribution. Next, we proposed a meta-learning episodic optimization scheme to adaptively learn a shared metric space between unseen samples and known attacks. Next, we developed a disentangled adversarial learning via an auxiliary batch norm to leverage both original training data and corresponding adversarial examples to train networks. Finally, the best-performing system updates the network parameters according to an integrated learning objective. Performance evaluation on the ASVspoof 2019 LA dataset confirms that our proposed approach effectively improves robustness/accuracy for spoofing detection system operation, which delivers results in terms of a pooled EER of 0.87\%, and a min t-DCF of 0.0277.

%\section*{Acknowledgment}
%The preferred spelling of the word ``acknowledgment'' in American English is 

\bibliographystyle{IEEEtran}
\bibliography{main.bib}

\begin{IEEEbiography}
[{\includegraphics[width=1in,height=1.25in,clip,keepaspectratio]{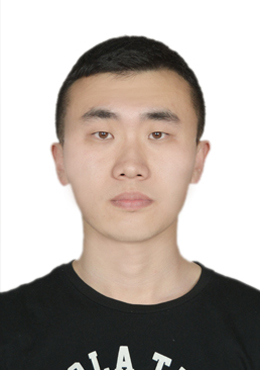}}]{Zhenyu Wang}
received the B.S. degree majoring in digital media technology from Hangzhou Dianzi University, Hangzhou, China in 2015. He received the M.S. degree in engineering computer system architecture from Beijing Language and Culture University, Beijing, China in 2019. He started working toward the Ph.D. degree in computer engineering at the University of Texas at Dallas (UTD), Richardson, TX, USA, in 2019. He works with Professor John H.L. Hansen at the Center for Robust Speech Systems.  Since the same year, he has been a Graduate Research Assistant with the Center for Robust Speech Systems (CRSS), UTD. His research interests include mispronunciation verification, computer-assisted language learning, forensic audio analysis and model adaptation for open-set speaker recognition system, representation learning used for acoustic modeling, keyword spotting, anti-spoofing, Audio Generation, LLM. He has authored over ten journal and conference papers in the field of speech processing and language technology. 
\end{IEEEbiography}

\begin{IEEEbiography}[{\includegraphics[width=1in,height=1.25in,clip,keepaspectratio]{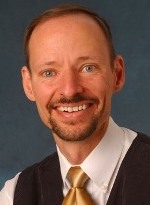}}]{John H. L. Hansen}
John H. L. Hansen (S’81–M’82–SM’93–F’07) received the B.S.E.E. degree from the College of Engineering, Rutgers University, New Brunswick, NJ, USA, in 1982, the M.S. and Ph.D. degrees in electrical engineering from Georgia Institute of Technology, Atlanta, GA, USA, in 1983 and 1988, respectively. He joined Erik Jonsson School of Engineering and Computer Science, University of Texas at Dallas (UTDallas), in 2005, where he currently serves as a Jonsson School Associate Dean for Research, as well as a Professor of Electrical and Computer Engineering, the Distinguished University Chair in Telecommunications Engineering, and a joint appointment as a Professor in the School of Behavioral and Brain Sciences (Speech \& Hearing). He previously served as a UTDallas Department Head of Electrical Engineering from August 2005 to December 2012, overseeing a +4× increase in research expenditures (4.5 to 22.3 M) with a 20\% increase in enrollment along with hiring 18 additional T/TT faculty, growing UTDallas to the eighth largest EE program from ASEE rankings in terms of degrees awarded. At UTDallas, he established the Center for Robust Speech Systems (CRSS). Previously, he served as a Department Chairman and Professor of Speech, Language and Hearing Sciences, and a Professor in Electrical and Computer Engineering, University of Colorado-Boulder (1998–2005), where he co-founded and served as an Associate Director of the Center for Spoken Language Research. In 1988, he established the Robust Speech Processing Laboratory and continued to direct research activities in CRSS at UTDallas. He is the author/coauthor of 800+ journal and conference papers including 13 textbooks in the field of speech processing and language technology, signal processing for vehicle systems, coauthor of textbook: Discrete-Time Processing of Speech Signals (IEEE Press, 2000), co-editor of DSP for In-Vehicle and Mobile Systems (Springer, 2004), Advances for In-Vehicle and Mobile Systems: Challenges for International Standards (Springer, 2006), In-Vehicle Corpus and Signal Processing for Driver Behavior (Springer, 2008), and lead author of the report The Impact of Speech Under Stress on Military Speech Technology, (NATO RTOTR-10, 2000). He has supervised 95 Ph.D./M.S. thesis candidates (54 Ph.D., 41 M.S./M.A.). His research interests include the areas of digital speech processing, analysis and modeling of speech and speaker traits, speech enhancement, feature estimation in noise, signal processing for hearing impaired/cochlear implants, robust speech recognition with emphasis on machine learning and knowledge extraction, and in-vehicle interactive systems for hands-free human-computer interaction. Dr. Hansen received the honorary degree Doctor Technices Honoris Causa from Aalborg University (Aalborg, DK) in April 2016, in recognition of his contributions to speech signal processing and speech/language/hearing sciences. He was recognized as an IEEE Fellow (2007) for contributions in “Robust Speech Recognition in Stress and Noise,” International Speech Communication Association (ISCA) Fellow (2010) for contributions on research for speech processing of signals under adverse conditions, and received The Acoustical Society of Americas 25 Year Award (2010) in recognition of his service, contributions, and membership to the Acoustical Society of America. He previously served as ISCA President (2017–2021) and Vice-president (2015-2017) and currently serves as tenure and member of the ISCA Board, having previously served as the Vice-President (2015–2017). He also is serving as a Vice-Chair on U.S. Office of Scientific Advisory Committees (OSAC) for OSAC-Speaker in the voice forensics domain (2015–2017). Previously, he served as an IEEE Technical Committee (TC) Chair and member of the IEEE Signal Processing Society: Speech-Language Processing Technical Committee (SLTC) (2005–2008; 2010–2014; elected IEEE SLTC Chairman for 2011–2013, Past-Chair for 2014), and elected as an ISCA Distinguished Lecturer (2011–2012). He has served as the member of the IEEE Signal Processing Society Educational Technical Committee (2005–2008; 2008–2010); Technical Advisor to the U.S. Delegate for NATO (IST/TG-01); IEEE Signal Processing Society Distinguished Lecturer (2005–2006), Associate Editor for IEEE TRANSACTION SPEECH AND AUDIO PROCESSING (1992–1999), Associate Editor for IEEE SIGNAL PROCESSING LETTERS (1998–2000), Editorial Board Member for IEEE SIGNAL PROCESSING MAGAZINE (2001–2003); and Guest Editor (October 1994) for special issue on Robust Speech Recognition for IEEE TRANSACTION SPEECH AND AUDIO PROCESSING. He is serving as an Associate Editor for JASA, and served on Speech Communications Technical Committee for Acoustical Society of America (2000–2003). He was the recipient of The 2005 University of Colorado Teacher Recognition Award as voted on by the student body. He organized and served as the General Chair for ISCA INTERSPEECH-2002, September 16–20, 2002, Co-Organizer, and Technical Program Chair for IEEE ICASSP-2010, Dallas, TX, USA, March 15–19, 2010, and Co-Chair and Organizer for IEEE SLT-2014, December 7–10, 2014 in Lake Tahoe, NV, USA.
\end{IEEEbiography}

\EOD

\end{document}